# The Missing Memristor:
# Novel Nanotechnology or rather new Case Study for the Philosophy and Sociology of Science?

Sascha Vongehr*,**,a

*National Laboratory of Solid-State Microstructures, Nanotechnology Group, Nanjing University, Hankou 22, Nanjing 210093, P. R. China
**Department of Philosophy, Nanjing University, Hankou 22, Nanjing 210093, P. R. China

In 2008, it was widely announced that the missing memristor, a basic two-terminal electrical circuit element, had finally been discovered. The memristor is the fourth and last such circuit element and thus completes circuit theory. Predicted already in 1971, the eventual discovery of something seemingly so basic needed almost 40 years. However, this discovery is doubted. The predicted memristor has no material memory and is based on magnetic flux, but the discovered devices constitute analogue memory storages that do not involve magnetism. The person who originally proposed the memristor did not reject the discovery but instead changed his mind about what a memristor is. We introduce the history and then memristance and the memristor as such. We discuss its status as a model rather than a device. We discuss the discovered devices, their stability, and how stability relates to the consistency of the theoretical entities. Also a thought experiment assuming a world without magnetism is presented. Inductors cannot exist there, but memory resistors could still be constructed. On the same grounds as the memristor was historically predicted, an "inductor" could then be predicted. Likely, somebody would also 'discover' one. A tentative sociological analysis compares to the flawed detection of gravitational waves but comes to very different conclusions.

---

[a] Corresponding author's electronic mail: vongehr@usc.edu

# 1 Introduction

The resistor, the capacitor, and the inductor are three well known, basic two-terminal circuit elements (B2TCE). In 1971, Leon Chua postulated a fourth, the memristor (Chua 1971)[1], on grounds of symmetry arguments, i.e. for the sake of completeness of circuit theory. The three well known ones correspond to extremely simple devices. Nevertheless, the suggested memristor was not discovered for almost 40 years, and it may still have not been discovered. In 2008, "The missing memristor found" (Strukov 2008)[2] was published in the respected science journal Nature and the memristor's supposed discovery was announced on the front pages of most major newspapers. This was indeed interesting news for basic physics, electrical engineering, and nanotechnology, the very field in which the devices in question were discovered, except for that the actually discovered devices had been discovered before 2008 already, yet few cared much (Chandra 2010)[3]. Nevertheless, the initiated theoretical advances in modeling memristic devices are considerable (Strukov 2008, Yang 2008, Strukov 2009)[2,4,5] and their potential applications in integrated circuits immense: They allow extended functionalities incorporating analog computation and self-programming neural networks and may thereby help approaching information processing the way human brains do it.

The memristor is also significant for the philosophy and sociology of science. This is due to its being predicted, its long absence, its discovery, and the (largely *missing*) controversy (Chandra 2010, Williams 2010)[3,6] around the latter. The eventual discovery of something that was naively expected to be maybe as simple as a capacitor or inductor, which can be constructed from just two metal plates or a metal coil, respectively, with a delay of almost 40 years is intriguing by itself. The long delay is perhaps a symptom of there being no such thing as a real memristor and indeed there are doubts about its discovery: The predicted memristor involves magnetic flux, but the discovered devices involve none. In fact, the discovered devices are similar to devices discovered in 1995 (Upadhyaya 1995)[7] and those early discoverers did and still do not think that their devices are memristors. As if this would not be enough, the following certainly submits the memristor as a case for the sociology of science: Leon Chua did not reject the purported discovery in 2008, but instead changed his mind about what a memristor is (Di Ventra 2009)[8]. One wonders why and particular suspicions are strengthened by the flood of papers that was generated and marketed under the catchy "memristor" label. Before one can embark on sociological analysis, there needs to be a rigorous critique of the discovery, which is the main aim of this work. If the memristor has not been found, its hyped discovery will only serve to hinder actually discovering the memristor or realizing that such is impossible as a device.

  Naively, the memristor issue could be added to success stories like Le Verrier's prediction of the existence and location of Neptune in 1846 and Dirac's prediction of the positron in 1928. Mendeleev's prediction of undiscovered chemical elements in 1870 rested on the still empty cells in the periodic table of the elements. The memristor is similarly a vacant spot in a two dimensional table (Fig. 1b), though a much smaller one. Given the symmetry we will discuss, somewhat similar is Murray Gell-Mann noticing missing pieces in the patterns of SU(3) symmetry representations, which led him to successfully predict subatomic particles. The mentioned historically predicted entities were discovered relatively soon after having been foretold. The memristor discovery reminds of the purported detection of gravitational waves by Joseph Weber in the late 1960s, which after an initial acceptance was discredited in the mid 1970s and subsequently led to a number of analyses (Collins [1985] 1992, Bartusiak 2000)[9,10]. More sensitive detectors could not reproduce Weber's claims. IBM physicist Richard Garwin built a similar detector but could only find one pulse in half a year, and that pulse was due to noise (Bartusiak 2000)[10]. Some rebuttals reviewed Weber's data analysis. The physicist David Douglass found an error in Weber's computer program which combined noise and artifices due to how the data was divided into batches which then resulted in daily coincidence signals (Levine 2004)[11]. The earliest rebuttals however were of a theoretical, even philosophical nature, similar to those we will apply to the memristor. For instance, Garwin pointed out that if Weber's detection were real, the universe would convert all of its energy into gravitational radiation in a matter of only 50 million years.

In both cases, the entity in question was predicted on theoretical grounds but may conceivably not verifiable. At the time of the first claimed detection of gravitational waves in 1969, their measurability was doubted even by those comfortable with general relativity: Would not the measure tape contract exactly along with the space contraction due to gravity waves? The originally proposed memristor may be impossible as a really existing device. Both times there are devices and real observations involved. As far as we know, Weber did not consciously make up any data; he just honestly reported what his visual system's acute pattern-recognition told him when confronted with noise in what is a naïve way of data analysis. Later on he reported what the flawed computer software calculated. However, the cases are also quite different. Gravitational waves have been inferred from convincing astrophysical observations, but they still have not been detected. If you ask for a memristor, you may be given an actual device and the claim

"here it is, you are holding one in your own hands". That object sitting in your palm is certainly not a misinterpretation of noise by software. If that object is not a memristor, something other than engineering proper must have gone astray.

Physics itself refused Weber's claims relatively fast. Particularly those differences that make the memristor issue sociologically interesting seem to ensure that science proper will not deliver speedily again without some outside prodding. Not just one Joseph Weber fears for his reputation, but a well funded scientific sub-field wants to believe into the memristor discovery. The main aim of this work is twofold: Firstly, the memristor debacle needs to be brought to the attention of a wider audience. Secondly, In order to justifiably submit the case, there must be a focused and solid argument on the grounds of which the case can be accepted as controversial in the first place.

We include a self contained introduction on what the purported discovery is and why it is questionable. We have reduced technicalities to the bare minimum, but rigorous argumentation cannot do entirely without them. Section 2 introduces memristance and the memristor. Also the discovered thin film devices will be described in a self contained fashion (Section 3). Explaining the memristor as a model of circuit theory (Section 4) and discussing its stability (Section 5) casts doubts on the necessity and viability of a memristor as an actual device. Section 6 assumes a world without magnetic fields. In this hypothetical world, the now newly discovered devices would have been discovered, too, but inductors could not exist. Identifying the devices as memristors would predict a missing inductor on the same grounds as the memristor was historically predicted. Section 7 suggest how a sociological analysis could start.

## 2 Carefully Introducing Memristance and the Memristor

### 2.1 Two fundamental relations and Basic 2-Terminal Circuit Elements

The memristor was predicted in the context of electrical circuit theory. Circuit theory has two *fundamental relations*, which derive from Maxwell's equations[b] via 'distilling' out the electric and the magnetic parts. The involved integrations of the free charge and current densities result in the charge $Q$ and current $I$, respectively. This leads to the *current-charge relation* that defines the current $I$ as the time derivative (written d/d$t$) of the electrical charge $Q$:

$$I = dQ/dt \qquad (1)$$

The integration of the magnetic field $B$ results[c] in the flux $\varphi$, while the electric field $E$ integrates to the induced voltage $U$. Flux $\varphi$ is not just some abstract variable. It is the *magnetic* flux and also called "flux-linkage" because it links the magnetic field to the magnetically induced voltage. Thus results the second fundamental relation, the *voltage-flux (or flux-linkage) relation*, which relates voltage $U$ to the time derivative of the flux $\varphi$:

$$U = d\varphi/dt \qquad (2)$$

One should be careful to not confuse the two fundamental *relations* with the two fundamental *laws* of circuit theory. The laws are: (1) The conservation of charge, which leads to Kirchhoff's node rule (all currents into and out of a circuit's network node sum to zero: $\Sigma I = 0$). (2) Energy conservation, which leads to Kirchhoff's loop rule (all voltages around any closed loop in the network sum to zero Σ $U = 0$). These two are not the source of the fundamental circuit theory relations. Energy and charge are also conserved in a hypothetical world without magnetic flux, where flux could be merely defined as the integration of a voltage over time consistent with Eq. (2). Leon Chua originally insisted on magnetism:

> "the physical mechanism characterizing a memristor device must come from the instantaneous (memoryless) interaction between the first-order electric field and the first-order magnetic field" (Chua 1971)[1].

---

[b] Maxwell's equations are the two pairs ($\nabla \cdot \vec{D} = \rho_{\text{free}}$ & $\nabla \times \vec{H} = \vec{j}_{\text{free}} + d\vec{D}/dt$) and ($\nabla \cdot \vec{B} = 0$ & $\nabla \times \vec{E} = -d\vec{B}/dt$). The free charge and current densities $\rho_{\text{free}}$ and $j_{\text{free}}$ occur in the first pair.

[c] $\varphi = -\int_{\text{Area}} d\vec{f} \times \vec{B}$

The two fundamental circuit theory relations are like two opposite edges of a tetrahedron (Fig. 1a), suspending the four fundamental circuit variables $I$, $Q$, $U$, and $\varphi$ at its corners. Eq. (1) provides $I$ and $Q$ while Eq. (2) holds $U$ and $\varphi$. This tetrahedral symmetry is what underlies the prediction of the memristor. There are four further edges of the tetrahedron: $Q$ to $U$, $U$ to $I$, $\varphi$ to $I$, and $\varphi$ to $Q$. These correspond to four further binary relations apart from the two fundamental relations.

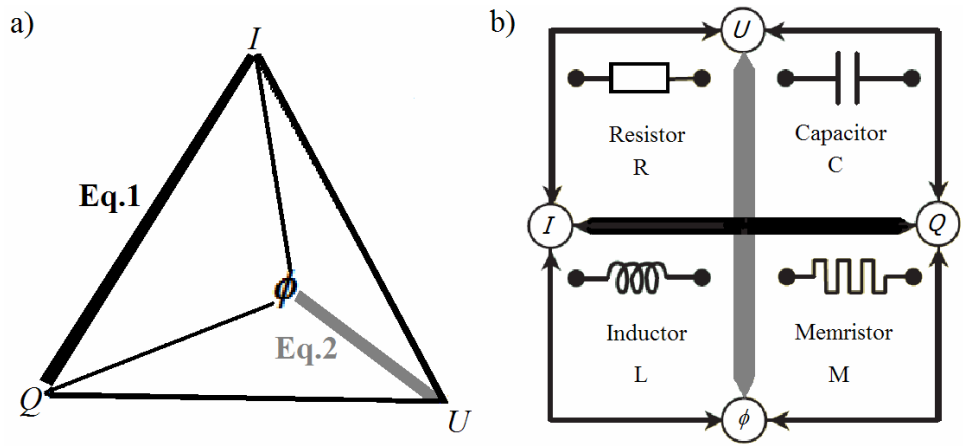

**Fig. 1** Illustrations of the symmetry that led to proposing the memristor: (a) the tetrahedron spanned by the four fundamental circuit variables, (b) the relations and circuit symbols of the four B2TCE that correspond to the four unlabeled edges in (a).

The four binary relations lead us directly to the basic two-terminal circuit elements (B2TCE), the first three of which are well known: The capacitor with capacitance $C_{(U,Q)} = dQ/dU$, the resistor with resistance

$$R_{(U,I)} = dU/dI,  \qquad (3)$$

and the inductor with inductance $L_{(\varphi,I)} = d\varphi/dI$. All of these are "basic" in the following sense: (1) As indicated by their symbols (Fig.1b), they have only two terminals – in and out, or plus and minus. (2) They are passive, meaning they do not supply any energy. A battery is therefore not a B2TCE. (3) The word "basic" also refers to that they are independent of each other, i.e. they span a space much like a *basis* of linearly independent vectors or axioms. For example: One cannot connect resistances and capacities together and end up with a circuit having inductance. Moreover, very simple devices are close analogies of these elements, which are as far as we introduced them, theoretical entities. The simplicity of existing devices is explicitly not what the term 'basic' implies.

### 2.2  *Memristance and Memristor: How the Prediction came to be*

The fourth B2TCE was predicted on grounds of symmetry. There is the as yet not discussed 'magnetic' edge of the tetrahedron which relates $\varphi$ and $Q$. This is the so called memristance

$$M_{(\varphi,Q)} = d\varphi/dQ. \qquad (4)$$

*M* is best understood by considering a purely charge dependent $M_{(Q)}$. Rewrite Eq. (4) as $d\varphi = M_{(Q)} dQ$ and integrate this over time. Eq. (1) and (2) tell us that the result is

$$U_{(t)} = M_{(Q_{(t)})} I_{(t)}. \tag{5}$$

Comparison with Eq. (3) shows that $M_{(Q)}$ is a resistance. Its units, i.e. the standard units of $\varphi$ divided by $Q$, are the same as that of resistance $R$, namely the Ohm ($\Omega$). The derivation also shows that this resistance depends on the charge $Q_{(t)}$. In other words, the resistance seems to "re*mem*ber" or "*mem*orize" the charge that has flowed through it; hence the term "memristance". However, keep the phrase "instantaneous (memoryless) interaction" in Chua's quote above in mind! There is no physical memory implied, but such is obviously present in the discovered devices.

Memristance $M$ closes circuit theory and exists by definition. The memristor is the corresponding B2TCE, just like the resistor is the element that has resistance. A resistor is commonly understood to be more than a symbol in a circuit diagram. Real resistor, capacitor, and inductor devices exist. Hence, there is no traditional sharp terminology that would distinguish the circuit theoretical entities from the real things. This leads to confusion about memristance and the corresponding memristor: It is the corresponding, perhaps even simple memristor *device* that has been postulated in 1971.

## 3   The Discovered Memristic Devices

The dependence of resistance on the charge that has flowed already is known from thin films in nanotechnology. This was discovered years before 2008, for example by some of the same authors who in 2008 claimed the memristor's discovery. Before 2008, the memristor analogy was merely not yet drawn upon (Lau 2004)[12]. See also others and references therein on memory resistance (Wu 2007)[13] and nonvolatile memory applications (Waser 2007)[14]. Memristic behavior is even common in semiconductor spintronic devices (Pershin 2008)[15].

In 2008, *the* memristor was announced (Strukov 2008)[2]. An introduction on thin film nanotechnology is beyond the scope of the present article, but the vital in light of our main topic can be understood easily. In one model, called the coupled ionic and electronic transport model (Strukov 2008, Yang 2008, Strukov 2009)[2,4,5], the current through the film consists partially of ionic charge carriers, for instance oxygen vacancies. These impurities enter the film (e.g. a titanium oxide junction) and thereby lead to a doping of the film with such impurities. This doping lowers the resistance of the film to other charge carriers, especially electrons. Within different memristic systems, it may not be ions but perhaps in the film induced metallic precipitates that lower the overall resistance of the film. However the case may be, as the width $w$ of the doped region grows, the resistance of the whole film decreases. The overall current and the impurity current are proportional to each other. Therefore, the overall resistance depends directly on the charge that has already passed through the film. This phenomenon has been discovered in thin films that are a few nanometers thick (say $D \sim 5$ nm), because the mobility of the impurities is small. The doped region grows slowly ($dw << D$) and the time over which the resistance changes is proportional to the film's thickness squared ($t \sim D^2$). Hence, the effect is not detectable in a macroscopic sample; a five micrometer instead of five nanometer thick sample will lead to an a million times smaller effect. The other reason for memristance being a thin film effect is the instability of the memristor as a device generally, as will be explained further below.

No magnetic fields are involved here. What reminds of the originally proposed memristor is that the thin film sandwiches are two-terminal elements and that their resistance changes according to the charge that has flown already. The thin film is a memory resistor, and so is any kludge that adjusts a resistance by tallying up the flown charge. The thin film's doped region is itself the physical memory that allows the device to be simple, but where is the "memoryless" memristor and its magnetism?

## 4   Theoretical Entities in Modeling versus Devices

The four B2TCE are circuit theoretical models rather than devices. Even ideal resistors do not exist. For instance, every actual capacitor also has some resistance. Ideal B2TCE are all impossible as devices. The following further example for the impossibility of ideal devices will demonstrate the use of the ideal theoretical entities: Connecting an ideal capacitor to an ideal voltage source would result immediately in an infinite current (for an infinitesimal amount of time). Thus, there must always be a resistor in there, too (Fig. 2). This resistor is not an actual device that we need to put in, but something that we put into the circuit diagram and into equations. It models the internal resistance of any actual voltage source like the battery.

The resistance $R$ implied by Eq. (3) models a part of a given circuit. It is less important that there actually is a simple device; a real resistor whose behavior closely resembles such an $R$. Memristance $M$ comes in when modeling circuits whose behavior cannot be fully modeled with help of only $L$, $R$, and $C$.

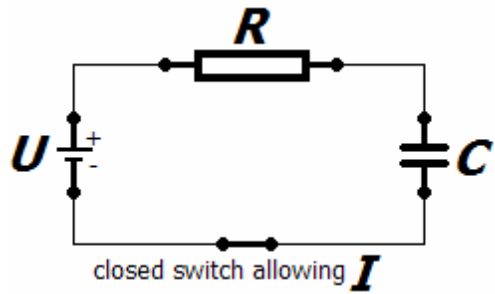

**Fig. 2** A voltage $U$ is connected to a capacitance $C$. There must be a resistance $R$, or otherwise there will be an infinite current $I$ drawn. This resistance is mostly due to the internal resistance of the battery. The existence of an actual resistor device is neither necessary nor implied.

Memristance $M$ can be exploited to model the thin film circuits during that special window in which the width $w$ of the doped region lies in between zero and the thickness $D$ of the thin film, i.e. $0 < w < D$ (as will be further discussed below). Such systems may be also modeled by charge dependent resistances $R_{(Q)}$, but this can be less elegant. Observing a memristic device in the current versus voltage plane, there will be anomalies like resistant switching, pinched hysteretic loops, hysteretic conductance as well as multiple conductance states, and apparent negative differential resistance, which have all been seen in thin films. Memristance $M$ is plotted in the $\varphi$ versus $Q$ plane, where such odd appearing anomalies can look surprisingly simple. Already in 1971, Chua remarks:

> "a memristor with a simple $\varphi$-$Q$ curve will give rise to a rather peculiar – if not complicated hysteretic – $U$-$I$ curve when erroneously traced in the current versus voltage plane. Perhaps, our perennial habit of tracing the $U$-$I$ curve of any new two-terminal device has already misled some of our device-oriented colleagues and prevented them from discovering the true essence of some new device…"[1].

He refers here to the proposed memristor, but it nevertheless underlines that the importance of the memristor concept is its usefulness in circuit theory, in device modeling, even if it does not exist as a physical object.

It should be remarked here that the dependence (on the involved fundamental variables) of any B2TCE is possibly non-linear. For example, the resistance $R_{(U,I)}$ may increase with the applied voltage. This is important, because a constant Ohmic resistor is a special case of a linear memristor [ $M = \mathrm{d}\varphi/\mathrm{d}Q = (\mathrm{d}\varphi/\mathrm{d}t)/(\mathrm{d}Q/\mathrm{d}t) = \mathrm{d}U/\mathrm{d}I$ ] and thus not independent. The point of needing all four B2TCE is that they are in general mutually independent. In the non-linear case, a resistance is neither a special case of a memristance, nor can it be modeled as a network of capacitances, inductances, and memristance.

Memristance is identical to resistance only in the linear case, but if $M$ is itself a function of charge, it is independent: A non-linear, passive memristic device cannot be made with networks of $L$, $C$, and $R$ without active components (batteries), and this is the significance of having memristance $M$. The discovered thin film sandwiches can be modeled via memristance only as long as the width $w$ of the doped region lies in between zero and the thickness $D$ of the thin film. This implies that the independence requirement of B2TCE is only satisfied if the electrical potential is artificially reversed whenever $w$ approaches the limits. The original proposal of the memristor envisioned an alternate-current (ac) device for the plain reason that electrical and magnetic fields were supposed to interact, but it did not artificially reverse current when some limit in physical memory has been exhausted. If such were allowed, any adjustable potentiometer regulated via a computer chip reading the current could count as a memristor.

That the independence requirement is only satisfied for limited (and practically short) times results in an intermediateness that we do not accept elsewhere. For example, the capacitor supplies energy when discharging and would not even count as passive if we were allowed to observe it for properly selected, short durations in order to categorize the device. Batteries are not capacitors and do explicitly not belong to the B2TCE, although they show similarities to de-charging capacitors. The latter is not a far fetched analogy given the necessary chemical (e.g. redox) reactions in the memristic thin film devices, which render the discovered devices indeed similar to batteries.

## 5   Consistency as a Model versus Stability of Devices

Two capacitances connected in parallel are plainly a higher capacitance: $C_{total} = C_{left} + C_{right}$. Similar equations hold for all four B2TCE. Regardless of whether they are connected in series, in parallel, or in more complicated networks, the network of multiple of any single particular type of the four (say twenty resistors) is in total just one such element. The relation between this consistency and the stability of the actual devices can be best comprehended with help of a simple plate capacitor. Such a capacitor is anyway a parallel arrangement of parts we can divide it into. In Fig. 2 for example, the left half of the capacitor is connected in parallel to the right half (this does not refer to the parallel arrangement of the two plates but to sawing the double plate along the vertical direction, resulting in two smaller double plates). There are unavoidable charge fluctuations in any real device, but a slightly more charged left half will not draw yet more charge from the right half. If such were the case, all charge would pile up in one tiny place on the capacitor, thereby violating consistency; $C_{total} = C_{left} + C_{right}$ would not be true. Such would result in the breakdown of capacitor devices. The piled up charge would overcome the resistance between the terminals' and discharge via a spark, something that usually destroys capacitors.

As just seen, theoretical consistency and device stability go hand in hand. The consistency of the memristor is ensured by circuit theory, but the discovered devices are unstable. They are thin films that are arranged between the terminals much like the plates of the capacitor in Fig. 2, meaning that the current does not go along the length of the film but through its width $D$. However, any part, say the left, that has by unavoidable charge fluctuation a somewhat larger $w_{left}$, will have less resistance and draw even more current, thereby increasing $w_{left}$ further, and so on. This runaway effect is another, unsurprisingly little advertised reason, for why these devices only work with extremely thin films. Thicker devices would suffer from that the low resistance region does not remain to be a very thin width $w$ along the whole device. Instead, soon after starting the device at $w = 0$, a filament breaks out from the low resistance region's front (Levine 2004, Lau 2004)[12,13].

The just discussed breakdown phenomenon destroys the assumed properties of any *single* memristor if it were imagined to be anything like the discovered devices! Capacitors and inductors can be connected in such a configuration that the network may be unstable and components are blown from the circuit board with alternate currents at resonance frequencies. However, every single capacitor stays to be a single capacitor and does not by itself fall apart after a short while, requiring to be described by a plethora of sub-capacitors in parallel. The discovered memristic devices commit such an atrocity. This casts another shadow over the interpretation of the thin film sandwiches as the originally proposed memristor. It moreover casts doubts on whether certain types of non linear memristors are possible at all as real devices. Although mathematically almost any odd resistance is allowed by Eq. (3), only quite simple non-linear resistors have actually been found.

## 6  Inductors in a World without Magnetism

Consider a world without magnetic fields. This is not too far fetched, since magnetic fields are a relativistic effect. If the velocity of light were much larger than it is already, moving charges would undergo much less Lorentz-Fitzgerald contraction and time dilatation so that we conceivably would have never discovered magnetism. In a world without magnetic fields, there would still be resistors and capacitors. There would be no principle hurdle in discovering the introduced thin film devices either. However, inductors and magnetic flux $\varphi$ are obviously absent.

According to the rather too convenient re-interpretations of flux (Di Ventra 2009)[8], one could then imagine that the scientists in that hypothetical world without magnetism would define $\varphi$ anyways as another useful variable, because the behavior of memory resistors can look simpler in the $\varphi$ versus $Q$ plane. Similar to the original, a 'non-magnetic L. Chua' postulates that there should be a fourth device, call it "inductor": there should be also a device that corresponds to the relation between current and flux. Ironically, the wave of papers triggered by the purported discovery of the memristor strongly suggests that also in the hypothetical world, "inductors" would be found and soon followed by announcements of "meminductors" (Di Ventra 2009)[8] and anything else imaginable. Nevertheless, we already know that in a world without magnetism, there are no magnetic fields, so a real inductor does strictly not exist.

For clarity, let us put this argument once more into a different setting: Imagine we could fully simulate a world without magnetism. Somebody living in that (virtual) world may invent some nanometer scale device and argue that the missing "inductor" has been finally found. Maybe the simulated scientific community would accept an inductor in a world without magnetic fields. Nevertheless, we secretly observe from the world that has magnetic fields. Do we not very well know that the found 'inductor' is a fake?

## 7 Starting the Sociological Analysis

In connection with the infamous flawed detection of gravitational waves by Joseph Weber in the late 1960s, there have been, after an initial acceptance of the results, strong rebuttals in the early 1970s. In the mid 1970s, the main stream ignored further articles by Weber; *Physical Review Letters* started to reject all his work. Collins referred to this late stage when he wrote (Collins 1999, abstract)[16]:

> "In physics, the literature is sufficiently open to allow some papers that have no credibility with the mainstream to be published. This normally causes no problem within 'core-groups' of scientists, because the orthodox interpretation is widely understood. There can still be trouble, however, from those who have not been socialized into the core-group's interpretative framework."

Collins defines the 'core-set' and 'core-group' so as to only include scientists actively working in the respective field. Outside of these groups, there are 'scientifically literate commentators', and further out still are policy makers and funding agencies, which is where the trouble comes in: Those outside cannot distinguish the faulty science. All papers that passed peer review look more or less equal, so policy makers can cherry pick according to their agenda. Funding resources get divided up, partially going into the wrong channels. The core-group understands certain claims to be flawed, which makes this analysis optimistic.

Collins' analysis applies to the gravitational wave case, but this will not be the right way to analyze the memristor case, where the core-group may not care about the faultiness of the memristor discovery. The difference is partially the size of the involved projects, since the memristor does not belong to 'Big Science' (1), and partially publish-or-perish culture (2) which nowadays overwhelms scientists with an unprecedented pressure to publish fast and plenty. Both render a strong rebuttal from inside of the involved scientific fields unlikely.

(1) Gravitational wave detectors are very large and expensive projects. Especially the newer detectors like the Laser Interferometer Gravitational-Wave Observatory (LIGO), which has two separate, four kilometer long arms, need collaborations of many scientists and non-scientific support staff. The above quote's "trouble" may arise and threaten a core-group's interest: Why fund their expensive project if competitors offer cheaper means that have perhaps already discovered the entity in question? Weber's resonance bars came basically for free in comparison with LIGO's initial costs being in excess of $300 million. The core-group's scientists have an obvious interest in attacking published yet flawed scientific articles.

Nanotechnology becomes one of the most important sciences, but there are by the nature of its subject matter no large and expensive projects like LIGO, the Large Hadron Collider (LHC) in particle physics, or the facilities trying to control nuclear fusion. The tacitly agreed upon consensus is that nanotechnology as such should get as much funding as possible and it does not matter much whether there is flawed science involved as long as the numerous small projects can go on and produce many small results that lead to further publications. There is no big issue that needs to be decided one way or another, nothing that the scientist is out to falsify, like say the Higgs boson in particle physics. It is questionable whether the concept of a core-group that has a 'socialized *interpretative framework*' is useful, since there are no big questions in need of interpretation, nothing comparable to gravitational waves that bend the very nature of space-time. The only big issue in materials science today is nanotechnology as such, which consists of a myriad of 'nano-projects', pun intended. Even any single researcher in the field of nanotechnology has usually numerous projects that need to have their funding ensured via the overall output of the respective laboratory, and not via any particular important result. These aims are best served with good news all around. All news about problems, say ethical or environmental concerns or internal fights in the community, are bad news that serve nothing and endanger the enthusiasm of the public in light of the great promises of nanotechnology. Critical papers in the field of nanotechnology (Vongehr 2010, 2011)[17,18] are extremely difficult to publish, with the degree of difficulty being only matched by the need for critical work in that very field.

Compared with other exact sciences, the reproducibility of results in nanotechnology and materials science is usually poor. Each particular result is of little importance and consequently, little value is given to reproduction. If something cannot be reproduced, it is more economical to explore something different fast, as the field is still wide open. It may be that one's own lab is just not able to reproduce a result for odd reasons, like a graduate student not working a centrifuge in the same wrong way as the one who was involved with the original work. Not that this is directly applicable to the memristor case, but it is beyond doubt that some scientific principles that were traditionally held in high esteem, especially reproducibility, are given little attention in these fields now.

(2) The large pressure exerted by publish-or-perish culture, which has never been as strong as today, contributes to corrupt research in the same direction as the dealt with smallish size of projects. Firstly, reproduction of another laboratory's results happens almost solely inherently, namely where new results are based on previous ones. Non-reproducible results do not get discredited but merely happen to never lead to any technology based on them. This holds for novel materials which do not actually exist or cannot be synthesized with the published methods, and for devices that do not work in the way they are described. This is also not directly applicable to the memristor, but the scientifically nonchalant culture that has developed in these fields has a deep impact.

In experimental high energy particle physics for example, the mere reproduction of discoveries like the top-quark are still important. In fields like nanotechnology, reproduction, on top of being uneconomical and unimportant, is dangerous in light of publish-or-perish. Firstly, articles are routinely rejected for not reporting novelties. So the researcher must decide: If I doubt a certain result, do I want to try and reproduce it? If the result is reproducible, it is not novel. Secondly, if I cannot reproduce it, publishing this negative result will be even more difficult. Every subfield is so specialized (admittedly, the smallness again) that peer reviewers will most likely include scientists loosely affiliated with the original researcher's claim. In the end, all I will effectively be able to proof is that I am not able to reproduce the result, say because of personal lack of ability or my laboratory's chemicals being impure. I will not have published enough this year and made some sworn enemies along the way. Reproduction has become a severe danger to the scientist's career. Publishing pressure affects all sciences, but the nature of materials science and nanotechnology's subject matter renders the impact in these fields extreme. Although super string theory is driven by publish-or-perish like few other fields, it is theoretical and the published papers' mathematical derivations can be checked. Data in experimental nanotechnology are taken on trust already between the principal investigator (PI) and her graduate student or post doctoral workers who perform the research, as the PI's role is that of an author and manager. For the scientists in the fields that were involved in the purported discovery of the memristor, publishing and reproducibility clash head-on, but publishing is all that matters. It is not even so that the players consciously avoid reproducing and there is certainly no conspiracy. A certain type of researcher is successful in this environment and shapes it further. A Darwinian co-evolution between systems and their environment turned too much of science into an almost pseudo-scientific, alchemist endeavor.

  In summary, the core-group has no interest in pointing out anything questionable with the discovery of the memristor, for instance as a competitor whose funding is at stake, and the culture is very different. It is not the culture known from earlier physics, where high level fights like the discussion between Einstein and Bohr increase the glamour. The new culture is largely formed by the pressure to publish fast which is detrimental to critical thinking. Whistle blowing is always and everywhere career suicide, but to criticize the memristor overtly does not even cross people's mind. In fact, those who criticized the purported memristor discovery did not doubt so much as instead complain that they did not get the same attention for their own, earlier devices. They felt left out. Deeper reaching criticism is perceived as treason or symptom of some deviance that just does not belong – it is "unprofessional". The memristor is such a catchy label and has proven so successful in the peer reviewed as well as popular media, all the researchers' interests are best served by jumping onto the bandwagon and somehow relating one's own ideas with the memristor. This strategy adds glamour to the otherwise mundane engineering devices, which are largely advertised by pointing out promising applications in medicine or data storage, but seldom allow entertaining the intrigue of fundamental symmetries.

Interpretive flexibility and cognitive as well as social interests are typical SSK categories. Some may hold that interpretive flexibility may play a central role. In fact, if one buys into social constructivism, the novel magnetism lacking interpretation of the memristor must be taken as a valid part of science on the grounds alone that at least publicly the majority of the involved scientists go happily along with it. There are many reservations against such positions and those reservations do not necessarily deny interpretive flexibility. There are fundamentally not decidable interpretative issues, as has been shown rigorously via so called 'dualities' inside of modern physics. However, we think the previous sections have convincingly shown that a non-magnetic memristor interpretation should not be accepted as a valid alternative. This leaves the social interests forming cognition. Modern publishing pressure must be taken into consideration when analyzing the sociological aspects surrounding the purported discovery of the memristor.

## 8   Conclusion

We have discussed several independent reasons to doubt the discovery of the memristor: among others, that the boundary conditions have to be artificially enforced or independence holds only intermediately and there is a further inherent instability in the devices. Distinguishing between models and devices clarified that apart from it behaving in a certain way, it is still not known what a memristor actually is, how we could build one. The 'basic' in B2TCE should not be misunderstood to mean as 'simple' as a capacitor. L. Chua originally insisted not only on the close relation between magnetism and the memristor but moreover on the interaction of both, the electric and the magnetic fields, making it a potentially very complex device:

> "an inductor has been identified to be an electromagnetic system where only the first-order magnetic field is negligible. […] The remaining case where both first-order fields are not negligible has been dismissed as having no corresponding situation in circuit theory. We will now offer the suggestion that this missing combination is precisely that which gives rise to the characterization of a memristor." (Chua 1971)[1]

The mutual independence of B2TCE ensures that one cannot connect many resistors, capacitors, and inductors in order to end up with a memristor. Components that behave similar to batteries are also not permitted. Something entirely new involving electric and magnetic fields is required. It is not enough to have something just somehow remembering resistance, say some ad hoc setup involving a computer. The memristor as differentiated from a broader class of memristic systems (Chua 1976)[19] is something that behaves according to Eq. (4), which involves magnetic flux. The memristor is after all proposed on grounds of the tetrahedral symmetry implied by the four fundamental circuit variables. We would not accept a capacitor without any charges $Q$, either.

Our section on the stability of memristor devices has questioned whether a memristor may exist as a stable device. Further research may prove the impossibility of such a device, justifying the usual attitude before 1971. Nevertheless, even if memristor devices are possible, the presence of physical memory in the discovered devices and especially the independent thought experiment cast strong doubts on whether the discovered devices are memristors. Our world has magnetic fields, and while Section 5 indicates that there are perhaps no real memristors at all, the non-magnetic world scenario doubts that a real memristor has been found: If it exists, it should look different. It should be something that does not lead to inconsistencies like inductors in a world without magnetism. The memristor is either impossible or still to be discovered.

Our comparison with Joseph Weber's claimed detection suggests that the memristor case is different and must ask how today's large pressure to publish in high impact journals highly impacts the quality of science and constructs knowledge. High impact citations and grants allotted to those who have them are basically all that counts in academic science today and this is disastrous in fields such as nanotechnology.